\documentclass[11pt,superscriptaddress,aps,prd,preprint]{revtex4}

\everymath{\displaystyle}
\usepackage{graphicx}

\usepackage{amsmath,amssymb,mathrsfs}

\newcommand{\bea}{\begin{eqnarray}}
\newcommand{\eea}{\end{eqnarray}}

\begin{document}

\title{The Role of Gravitational Energy Flux in Cosmic Acceleration}

\author{S. C. Ulhoa}\email[]{sc.ulhoa@gmail.com}
\affiliation{Instituto de F\'isica, Universidade de Bras\'ilia, 70910-900, Bras\'ilia, DF, Brazil} \affiliation{Canadian Quantum Research Center,\\ 
204-3002 32 Ave Vernon, BC V1T 2L7  Canada} 

\author{F. L. Carneiro}\email[]{fernandolessa45@gmail.com}
\affiliation{Universidade Federal do Norte do Tocantins,
Centro de Ci\^encias Integradas,
77824-838 Aragua\'ina, TO, Brazil}

\author{J. W. Maluf}\email[]{jwmaluf@gmail.com}
\affiliation{Instituto de F\'isica, Universidade de Bras\'ilia, 70910-900, Bras\'ilia, DF, Brazil} 


\begin{abstract}
The article deals with the role of gravitational radiation energy in the large-scale dynamics of the universe. Motivated by the observed accelerated expansion, we investigate whether gravitational energy, treated as a well-defined physical quantity within the teleparallel equivalent of general relativity, contributes to cosmological acceleration through its associated energy flux. Using radiative space-times described by the Bondi--Sachs framework, we analyze the total gravitational energy and the corresponding energy flux evaluated in asymptotic regions. Particular emphasis is placed on the cumulative character of gravitational radiation over long time scales and on the fact that gravitational energy in this formulation is not positively definite. The present analysis provides a consistent theoretical basis for assessing the relevance of gravitational radiation energy and its flux in cosmological contexts.
\end{abstract}

\maketitle

\date{\today}
\section{Introduction} \label{sec.1}

Since the independent measurements of the accelerated expansion of the universe by Riess \emph{et al.} and Perlmutter \emph{et al.} in their seminal works at the end of the 1990s~\cite{escura,perlmutter}, a physical explanation for this phenomenon has been actively sought. This effort characterizes the so-called dark energy problem, which remains open despite the various hypotheses proposed over the past decades, such as the introduction of a cosmological constant or dynamical fields of the quintessence type~\cite{lambda,phi}. More recently, new observational measurements have challenged the results obtained at the end of the twentieth century, reviving the debate on the interpretation of cosmic acceleration~\cite{acc, DESI2025}. In particular, the DESI collaboration has recently reported observational evidence for a time-evolving dark energy equation of state, challenging the standard $\lambda$CDM model~\cite{DESI2025}. The problem gains further relevance from the fact that none of the currently accepted hypotheses is able to consistently account for the apparent tensions among different observational indicators, such as the Hubble constant tension~\cite{tensaoH, H0DN2026}.

From an observational standpoint, the central problem is the accelerated expansion itself. This fact motivates the question of whether such observations may also serve as a test for the physical role of gravitational energy. The difficulty in defining gravitational energy is not new and arises already in the context of general relativity, where its localization is inherently problematic. A consistent and well-defined description of gravitational energy is instead achieved within the teleparallel equivalent of general relativity (TEGR), where gravitational energy and energy flux emerge as genuine physical quantities~\cite{maluf}. In this framework, gravitational energy is not introduced in a generic or phenomenological manner, but arises naturally from well-defined dynamical processes. In particular, gravitational radiation generated through radiative mass--energy processes, as described by the Bondi--Sachs metric~\cite{bondi1960, bondi,sachs}, provides a concrete mechanism for the production of gravitational energy. The accumulation of such radiation throughout cosmic history may lead to a gravitational radiation background that permeates space and interacts dynamically with matter. Owing to its effectively nonhomogeneous distribution, this background plays a dynamical role not only in the observed accelerated expansion of the universe, but also in the current tension associated with the Hubble constant. It is worth noting that gravitational energy in the TEGR framework is not positively definite in general. In fact, there exist physically relevant situations in which negative gravitational energy densities arise, a feature that becomes particularly prominent in the presence of gravitational waves, as reviewed in Ref.~\cite{maluf2023}. Gravitational radiation exhibiting this characteristic may propagate toward asymptotic regions without dissipation, allowing its associated energy flux to persist over large distances and long time scales.

The gravitational radiation is not expected to be an isolated phenomenon, but rather the result of frequent and cumulative processes occurring throughout cosmic history. As a consequence, the energy associated with gravitational radiation tends to build up over time (in an averaged sense) in regions far from the sources, progressively contributing to the large-scale gravitational environment. In this sense, an accelerated expansion driven by gravitational radiation energy is naturally expected to manifest itself as a large-scale effect. The contribution of a single radiative event would be negligible for an observer, but the cumulative effect of gravitational radiation produced over billions of years may become dynamically relevant at cosmological scales.  The purpose of this work is to analyze the total gravitational radiation energy-momentum flux in radiative Bondi-Sachs space-times and to determine its dynamical implications. While in Ref.~\cite{ulhoa2024} the energy of the gravitational radiation generated by the Bondi mechanism was investigated in connection with its possible manifestation as instrumental noise in gravitational-wave detectors, the present work addresses a distinct physical regime, focusing on the combined analysis of the energy flux and the momentum flux, the latter defining a radiative force. In particular, we show that both the energy and momentum fluxes associated with radiative space-times assume negative values, a feature that allows gravitational radiation to effectively transfer energy and momentum to the surrounding space-time and potentially drive accelerated expansion at cosmological scales.

This paper is organized as follows. In Sec.~\ref{sec.2}, the teleparallel formulation of gravity is briefly reviewed and the Bondi--Sachs radiative space-time is introduced as the framework for the analysis of gravitational energy and energy flux. In Sec.~\ref{sec.3}, we obtain and discuss the behavior of the total gravitational energy and energy flux in regions far from the radiating sources. Finally, in Sec.~\ref{sec.4}, our final remarks are presented.

\section{Teleparallel Formulation and Bondi--Sachs Radiative Space--Time} \label{sec.2}

\subsection{Teleparallelism Equivalent to General Relativity (TEGR)}

In this subsection, we provide a brief overview of TEGR, as extensively reviewed in Ref.~\cite{maluf}. The teleparallel equivalent of general relativity is an alternative formulation expressed in terms of tetrad fields, $e^{a}\,_{\lambda}$, yet dynamically equivalent to general relativity. The tetrad field naturally encodes two types of symmetries: Lorentz symmetry, represented by Latin indices, and diffeomorphism invariance, represented by Greek indices. Consequently, the tetrad field projects tensor components onto each of these symmetry sectors.

The tetrad field is related to the metric tensor through
\begin{equation}
g^{\mu\nu} = e^{a\mu} e_{a}\,^{\nu}\,,
\end{equation}
which implies that there is an infinite number of tetrads corresponding to a given metric tensor. This freedom in the choice of tetrads is intimately related to the selection of a reference frame. In other words, choosing a tetrad field is equivalent to choosing a frame adapted to a particular observer.

Although general relativity is usually formulated in terms of Riemannian geometry, the same dynamics can also be described using Weitzenb\"ock geometry. This equivalence arises from a well-defined mathematical identity between the connections associated with each geometry. Consider a space endowed with the following connection:
\begin{equation}
\Gamma_{\mu\lambda\nu} = e^{a}\,_{\mu}\, \partial_{\lambda} e_{a\nu}\,,
\end{equation}
whose torsion is given by
\begin{equation}
T^{a}\,_{\lambda\nu} = \partial_{\lambda} e^{a}\,_{\nu} - \partial_{\nu} e^{a}\,_{\lambda}\,. \label{4}
\end{equation}

If the Riemannian space is endowed with the Christoffel symbols ${}^0\Gamma_{\mu\lambda\nu}$, then the two connections are related by the identity
\begin{equation}
\Gamma_{\mu\lambda\nu} = {}^0\Gamma_{\mu\lambda\nu} + K_{\mu\lambda\nu}\,, \label{2}
\end{equation}
where
\begin{equation}
K_{\mu\lambda\nu} = \frac{1}{2} \left( T_{\lambda\mu\nu} + T_{\nu\lambda\mu} + T_{\mu\lambda\nu} \right) \,, \label{3}
\end{equation}
is the contortion tensor. It is important to note that the two geometries maintain a symmetric relationship. While the Riemannian geometry is characterized by a space with nonzero curvature and vanishing torsion, the Weitzenb\"ock geometry has nonzero torsion and vanishing curvature. These concepts should not be confused, as each dynamical quantity is computed from the corresponding connection associated with its respective geometry.

Since the curvature constructed from the Christoffel symbols is identically zero, it follows that
\begin{equation}
eR(e) \equiv -\, e\Big(\frac{1}{4} T^{abc} T_{abc} + \frac{1}{2} T^{abc} T_{bac} - T^a T_a \Big) + 2 \partial_\mu (e T^\mu)\,, \label{eq5}
\end{equation}
where $T^\mu = T^{\nu\mu}{}_\nu$. This allows us to define the Lagrangian density as
\begin{eqnarray}
\mathfrak{L}(e_{a\mu}) &=& -\kappa\, e \Big( \frac{1}{4} T^{abc} T_{abc} + \frac{1}{2} T^{abc} T_{bac} - T^a T_a \Big) - \mathfrak{L}_M \nonumber \\
&\equiv& -\kappa\, e \, \Sigma^{abc} T_{abc} - \mathfrak{L}_M\,, \label{6}
\end{eqnarray}
where
\begin{equation}
\Sigma^{abc} = \frac{1}{4} (T^{abc} + T^{bac} - T^{cab}) + \frac{1}{2} (\eta^{ac} T^b - \eta^{ab} T^c)\,, \label{7}
\end{equation}
is sometimes referenced as the superpotential  describing the interaction with matter fields $\mathfrak{L}_M$ in TEGR. The coupling constant is $\kappa = 1/(16\pi)$ in natural units with $G = c = 1$. It is also worth noting that the total divergence term in identity~\eqref{eq5} has been discarded in the Lagrangian density, as it does not contribute to the field equations.

The field equations derived from expression~\eqref{6} read
\begin{equation}
\partial_\nu \left( e \, \Sigma^{a\lambda\nu} \right) = \frac{1}{4\kappa} \, e \, e^a{}_\mu \left( t^{\lambda\mu} + T^{\lambda\mu} \right)\,, \label{10}
\end{equation}
where
\begin{equation}
t^{\lambda\mu} = \kappa \left[ 4 \, \Sigma^{bc\lambda} T_{bc}{}^\mu - g^{\lambda\mu} \, \Sigma^{abc} T_{abc} \right]\,, \label{11}
\end{equation}
represents the gravitational energy-momentum tensor. Since
\begin{equation}
\partial_\lambda \partial_\nu \left( e \, \Sigma^{a\lambda\nu} \right) \equiv 0\,, \label{12}
\end{equation}
a conserved quantity immediately follows. That is,
\begin{equation}
P^a = \int_V d^3x \, e \, e^a{}_\mu \left( t^{0\mu} + T^{0\mu} \right)\,, \label{14}
\end{equation}
which represents the total energy-momentum vector in this formulation. This conserved quantity can also be expressed in terms of $\Sigma^{a\mu\nu}$ as
\begin{equation}
P^a = 4\kappa \int_V d^3x \, \partial_\nu \left( e \, \Sigma^{a0\nu} \right)\,. \label{14.1}
\end{equation}
Accordingly, the total energy-momentum vector can be written as a surface integral. It is worth noting that $t^{\mu\nu}$ is a tensor under coordinate transformations, and that $P^a$ is covariant under Lorentz transformations. In other words, the energy-momentum vector does not depend on the choice of coordinates but transforms consistently under changes of frame. This property is not only essential for the definition of a meaningful energy-momentum vector but also a necessary condition for its very existence. In special relativity, the energy of a particle at rest is $mc^2$, independent of the coordinate system, while the energy of a moving particle is $\gamma mc^2$, reflecting its status as a component of the four-momentum. There is no reason to abandon these fundamental features for gravitational energy when matter interacts with the gravitational field.

From the conservation identity~\eqref{12}, one can expand the sum in the free index $\lambda$ to separate temporal and spatial contributions,
\begin{equation}
\partial_0 \partial_\nu ( e \, \Sigma^{a0\nu} ) + \partial_i \partial_\nu ( e \, \Sigma^{ai\nu} ) = 0.
\end{equation}
Noting that $\partial_\nu ( e \, \Sigma^{a0\nu} )$ is precisely the integrand of the total energy-momentum vector defined in~\eqref{14.1}, we can pass the spatial term to the right-hand side and integrate over a spatial volume $V$ to define the total gravitational energy-momentum flux
\begin{equation}
\frac{d P^a}{dx^0} = - \int_V d^3x \, \partial_i \partial_\nu ( e \, \Sigma^{ai\nu} ).
\end{equation}
Finally, applying Gauss's theorem, this expression can be rewritten as a surface integral over the boundary $S = \partial V$,
\begin{equation}
\frac{d P^a}{dx^0} = - \oint_S dS_i \, \partial_\nu ( e \, \Sigma^{ai\nu} ),
\end{equation}
providing a natural and rigorous definition of the total energy-momentum flux in terms of $\Sigma^{a\mu\nu}$. This construction includes contributions from both the gravitational field and matter, and arises directly from the fundamental conservation law of TEGR.


\subsection{Bondi--Sachs Radiative Space--Time}

In this subsection we describe the main results already presented in Ref.~\cite{ulhoa2024}. The Bondi--Sachs space-time, as formulated in the original works~\cite{bondi1960, bondi,sachs}, describes how mass is converted into gravitational radiation, that is, it characterizes a purely gravitational radiative configuration. The line element is written in terms of the coordinates $(u,r,\theta,\phi)$ as
\begin{eqnarray}
ds^2 &=& g_{00}\,du^2 + g_{22}\,d\theta^2 + g_{33}\,d\phi^2 \nonumber \\
&{}& + 2g_{01}\,du\,dr + 2g_{02}\,du\,d\theta + 2g_{03}\,du\,d\phi
+ 2g_{23}\,d\theta\,d\phi \,,
\label{9a}
\end{eqnarray}
where
\begin{eqnarray}
g_{00} &=& \frac{V}{r}e^{2\beta}
- r^2\!\left(e^{2\gamma}U^2\cosh 2\delta
+ e^{-2\gamma}W^2\cosh 2\delta
+ 2UW\sinh 2\delta\right), \nonumber \\
g_{01} &=& -e^{2\beta}, \nonumber \\
g_{02} &=& -r^2\!\left(e^{2\gamma}U\cosh 2\delta
+ W\sinh 2\delta\right), \nonumber \\
g_{03} &=& -r^2\sin\theta\!\left(e^{-2\gamma}W\cosh 2\delta
+ U\sinh 2\delta\right), \nonumber \\
g_{22} &=& r^2 e^{2\gamma}\cosh 2\delta, \nonumber \\
g_{33} &=& r^2 e^{-2\gamma}\cosh 2\delta\,\sin^2\theta, \nonumber \\
g_{23} &=& r^2\sinh 2\delta\,\sin\theta .
\label{10a}
\end{eqnarray}
The metric functions above admit the following asymptotic behavior:
\begin{eqnarray}
V &\simeq& -r + 2M, \nonumber \\
\beta &\simeq& -\frac{c^2 + d^2}{4r^2}, \nonumber \\
\gamma &\simeq& \frac{c}{r}, \nonumber \\
\delta &\simeq& \frac{d}{r}, \nonumber \\
U &\simeq& -\frac{l(u,\theta,\phi)}{r^2}, \nonumber \\
W &\simeq& -\frac{\bar l(u,\theta,\phi)}{r^2},
\label{11a}
\end{eqnarray}
with $$l=\partial_2 c+2c\,\cot\theta+\partial_3 d\,\csc \theta\,,$$
$$\bar{l}=\partial_2 d +2d\,\cot\theta -\partial_3 c\csc \theta\,.$$
Thus, the gravitational dynamics of the radiative space-time is encoded in
the mass aspect $M(u,\theta,\phi)$ and in the functions $c(u,\theta,\phi)$
and $d(u,\theta,\phi)$, whose time derivatives define the News functions
associated with gravitational radiation. The evolution of the mass aspect
is governed by
\begin{equation}
\partial_0 M
= -\bigl[(\partial_0 c)^2 + (\partial_0 d)^2\bigr]
+ \frac{1}{2}\,\partial_0\!\left(\partial_2 l
+ l \cot\theta
+ \frac{\partial_3 \bar l}{\sin\theta}\right),
\label{17a}
\end{equation}
which determines the behavior of $M$ in the line element. In the original
Bondi--Sachs formulation, the mass aspect provided the primary description
of the gravitational radiation emitted by the source. More recently,
however, it has been shown that the energy associated with the gravitational
radiation itself contributes in the asymptotic region, with this
contribution being expressed in terms of the News functions.

In order to obtain the total energy of the Bondi--Sachs space-time, it is
necessary to fix an observer. We choose a stationary observer by adopting
the tetrad field
\begin{equation}
e_{a\mu}=\scalebox{1.25}{$\left(\begin{smallmatrix}
-\mathcal{A} & -\mathcal{E} & -\mathcal{F} & -\mathcal{G} \\
0 & \mathcal{B}_2 \cos\theta \cos\phi + \mathcal{B}_1 \cos\phi \sin\theta - \mathcal{B}_3 \sin\theta \sin\phi
& r \mathcal{C}_1 \cos\theta \cos\phi - \mathcal{C}_2 \sin\theta \sin\phi
& -r \mathcal{D} \sin\theta \sin\phi \\
0 & \mathcal{B}_3 \cos\phi \sin\theta + \mathcal{B}_2 \cos\theta \sin\phi + \mathcal{B}_1 \sin\theta \sin\phi
& -\mathcal{C}_2 \cos\phi \sin\theta + r \mathcal{C}_1 \cos\theta \sin\phi
& r \mathcal{D} \cos\phi \sin\theta \\
0 & \mathcal{B}_1 \cos\theta - \mathcal{B}_2 \sin\phi
& -r \mathcal{C}_1 \cos\theta
& 0
\end{smallmatrix}\right)$}\,,
\end{equation}
where
\begin{eqnarray}
\mathcal{A} &=& 1 - \frac{M}{r}\,, \nonumber \\
\mathcal{E} &=& 1 + \frac{M}{r}\,, \nonumber \\
\mathcal{F} &=& -l - \frac{1}{r}\,(2cl + 2d\bar l + M l - p)\,, \nonumber \\
\mathcal{G} &=& -\sin\theta\left[\bar l + \frac{1}{r}\,(-2c\bar l + 2dl + M\bar l - \bar p)\right], \nonumber \\
\mathcal{B}_1 &=& 1 + \frac{M}{r}\,, \nonumber \\
\mathcal{B}_2 &=& -\frac{1}{r} - \frac{1}{r^2}\,(2Ml + cl - p)\,, \nonumber \\
\mathcal{B}_3 &=& -\frac{1}{\sin\theta}\left[\frac{\bar l}{r} + \frac{1}{r^2}\,(2M\bar l - c\bar l + 2dl - \bar p)\right], \nonumber \\
\mathcal{C}_1 &=& 1 + \frac{c}{r} + \frac{1}{r^2}\left(\frac{l^2}{2} + c^2 - d^2\right), \nonumber \\
\mathcal{C}_2 &=& \frac{1}{\sin\theta}\left[2d + \frac{1}{r}\,(l\bar l + 2cd)\right], \nonumber \\
\mathcal{D} &=& 1 - \frac{c}{r} + \frac{1}{r^2}\left(\frac{\bar l^2}{2} + c^2 + d^2\right). \nonumber
\end{eqnarray}
It is important to note that the functions $p$ and $\bar p$ do not appear in
the expression for the total energy; they are included here only for
completeness and are defined in Ref.~\cite{huang2006}. With this choice, the
total energy reads
\begin{equation}
P^{(0)}=4k\int_0^{2\pi} d\phi \int_0^{\pi} d\theta \,\sin\theta\,
\bigl[M+\partial_0 F\bigr],
\label{14a}
\end{equation}
where
\begin{equation}
F=-\frac{1}{4}\,\bigl(l^2+\bar l^2\bigr)+\frac{1}{2}\,c^2+d^2.
\label{15a}
\end{equation}
We emphasize that the term involving $F$ represents the contribution of the
gravitational radiation energy to the total energy in the asymptotic region.

We recall that $P^{(a)}$ is the energy-momentum \textit{vector} 
of the gravitational field, obtained by integration of the corresponding 
tensor density over a three-dimensional spatial volume. Its temporal 
component $P^{(0)}$ is therefore the total gravitational energy, while 
the spatial components $P^{(i)}$, presented below, correspond to the 
total linear momentum carried by gravitational radiation.

The spatial components of the Bondi four--momentum are given by
\begin{align}
P^{(1)} &= 4k \int_0^{2\pi} d\phi \int_0^{\pi} d\theta \Biggl\{
\sin^2\theta \cos\phi\,(\partial_0 F)
\nonumber\\
&\qquad
+ \frac{1}{4}\Bigl[
\sin\theta \cos\theta \cos\phi \bigl(2\partial_2 M + l\,\partial_0 M\bigr)
- \sin\phi \bigl(2\partial_3 M + \bar l \sin\theta\,\partial_0 M\bigr)
\Bigr]
\Biggr\},
\label{14b}
\\[1ex]
P^{(2)} &= 4k \int_0^{2\pi} d\phi \int_0^{\pi} d\theta \Biggl\{
\sin^2\theta \sin\phi\,(\partial_0 F)
\nonumber\\
&\qquad
+ \frac{1}{4}\Bigl[
\sin\theta \cos\theta \sin\phi \bigl(2\partial_2 M + l\,\partial_0 M\bigr)
+ \cos\phi \bigl(2\partial_3 M + \bar l \sin\theta\,\partial_0 M\bigr)
\Bigr]
\Biggr\},
\label{14c}
\\[1ex]
P^{(3)} &= 4k \int_0^{2\pi} d\phi \int_0^{\pi} d\theta \Biggl\{
\sin\theta \cos\theta\,(\partial_0 F)
- \frac{1}{4}\sin^2\theta \bigl(2\partial_2 M + l\,\partial_0 M\bigr)
\Biggr\}.
\label{14d}
\end{align}
The spatial components above quantify the linear momentum carried by gravitational radiation in each direction, enabling the analysis of the directional (anisotropic) effect of the radiative process on the asymptotic spacetime

\section{Total Gravitational Energy and Energy Flux in Radiative Space--Times} \label{sec.3}

In this section we present the numerical results associated with radiative Bondi--Sachs space-times, focusing on the behavior of the total gravitational energy and its corresponding energy flux for specific choices of the radiative functions $c(u,\theta,\phi)$ and $d(u,\theta,\phi)$. Our analysis is based on explicit mode decompositions of the radiative degrees of freedom, first in terms of spherical harmonics and subsequently in more restricted configurations, allowing us to examine how the gravitational energy flux depends on the angular structure of the radiation. The results are presented in the form of time-dependent energy and flux profiles, obtained under well-defined initial conditions, which correspond to an effective transfer of gravitational energy associated with localized radiative processes.

As a first step, we consider a general radiative configuration in which the angular dependence of
the Bondi--Sachs degrees of freedom is described by a truncated expansion in spherical harmonics.
The functions $c$ and $d$ are written as
\begin{equation}
c(u,\theta,\phi) =
\sum_{\ell=\ell_{\min}}^{\ell_{\max}}
\left[
a_{\ell 0}(u)\,\mathcal{Y}_{\ell 0}(\theta,\phi)
+ 2 \sum_{m=1}^{\ell}
a_{\ell m}(u)\,\mathrm{Re}\!\left(\mathcal{Y}_{\ell m}(\theta,\phi)\right)
\right],
\end{equation}
with an analogous expression for $d(u,\theta,\phi)$ and
\begin{equation}
\mathcal{Y}_{\ell m}(\theta,\phi)\equiv
\sin\theta\,\partial_{\theta}\!\left(\sin\theta\,\partial_{\theta}Y_{\ell m}(\theta,\phi)\right)\,,
\end{equation}
where $Y_{\ell m}$ are the Spherical Harmonics. This real representation ensures that all
physical quantities remain explicitly real. The time dependence of the modes is encoded in smooth
coefficients $a_{\ell m}(u)$, chosen here as Gaussian functions of the retarded time,
\begin{equation}
a_{\ell m}(u) = \frac{1}{(\ell+m)!}\, e^{-u^{2}},
\end{equation}
which describe localized bursts of gravitational radiation. In all cases considered here, the Bondi mass aspect is not prescribed a priori, but dynamically
evolved by numerical integration of the Bondi--Sachs evolution equation,
Eq.~(\ref{17a}), using the chosen radiative functions $c(u,\theta,\phi)$ and $d(u,\theta,\phi)$ as inputs.
Starting from a constant initial mass aspect $M_{0}$, this procedure ensures that the total
gravitational energy and momentum displayed in the figures consistently incorporate the full
radiative backreaction encoded in the Bondi formalism.

\begin{figure}[t]
\centering
\begin{minipage}{0.48\linewidth}
\centering
\includegraphics[width=\linewidth]{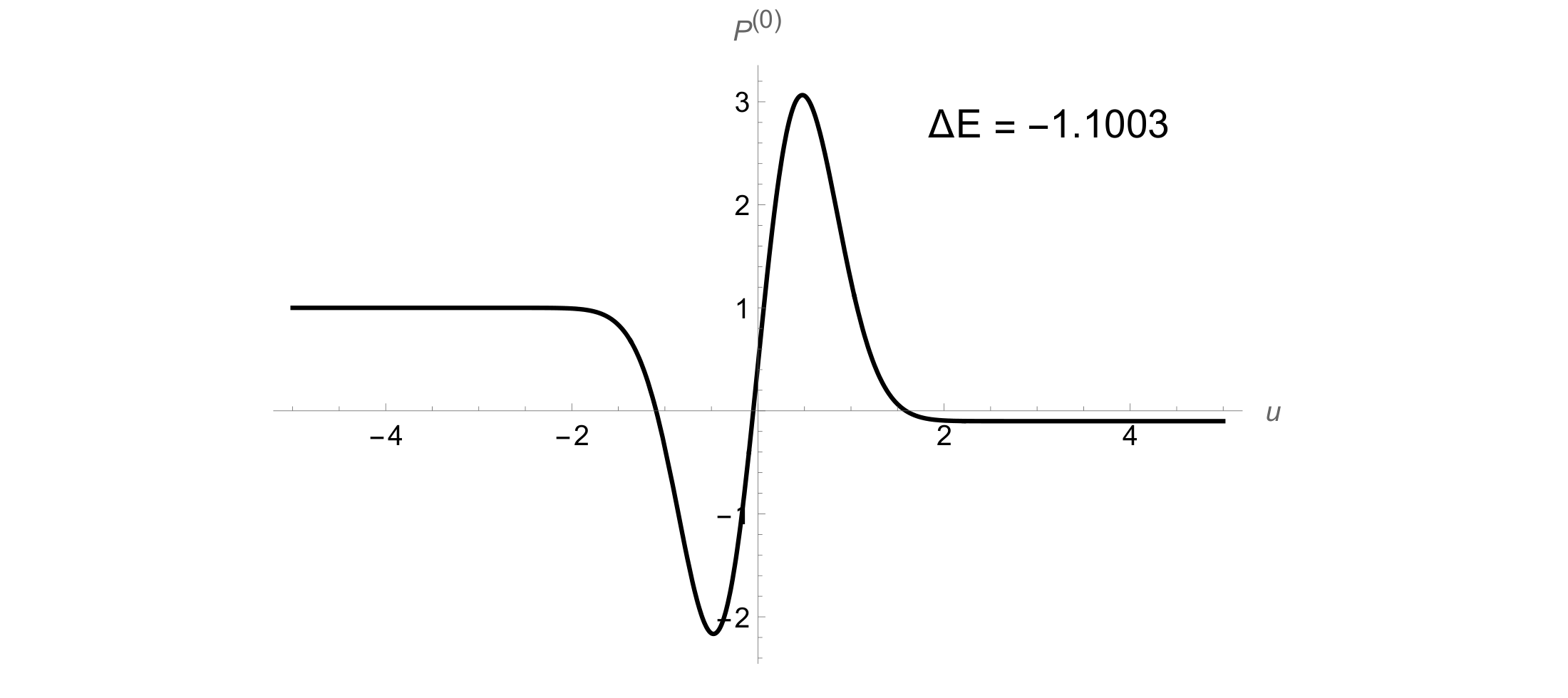}
\par\smallskip
{\small (a) Total Bondi-Sachs gravitational energy $P^{(0)}(u)$.}
\end{minipage}\hfill
\begin{minipage}{0.48\linewidth}
\centering
\includegraphics[width=\linewidth]{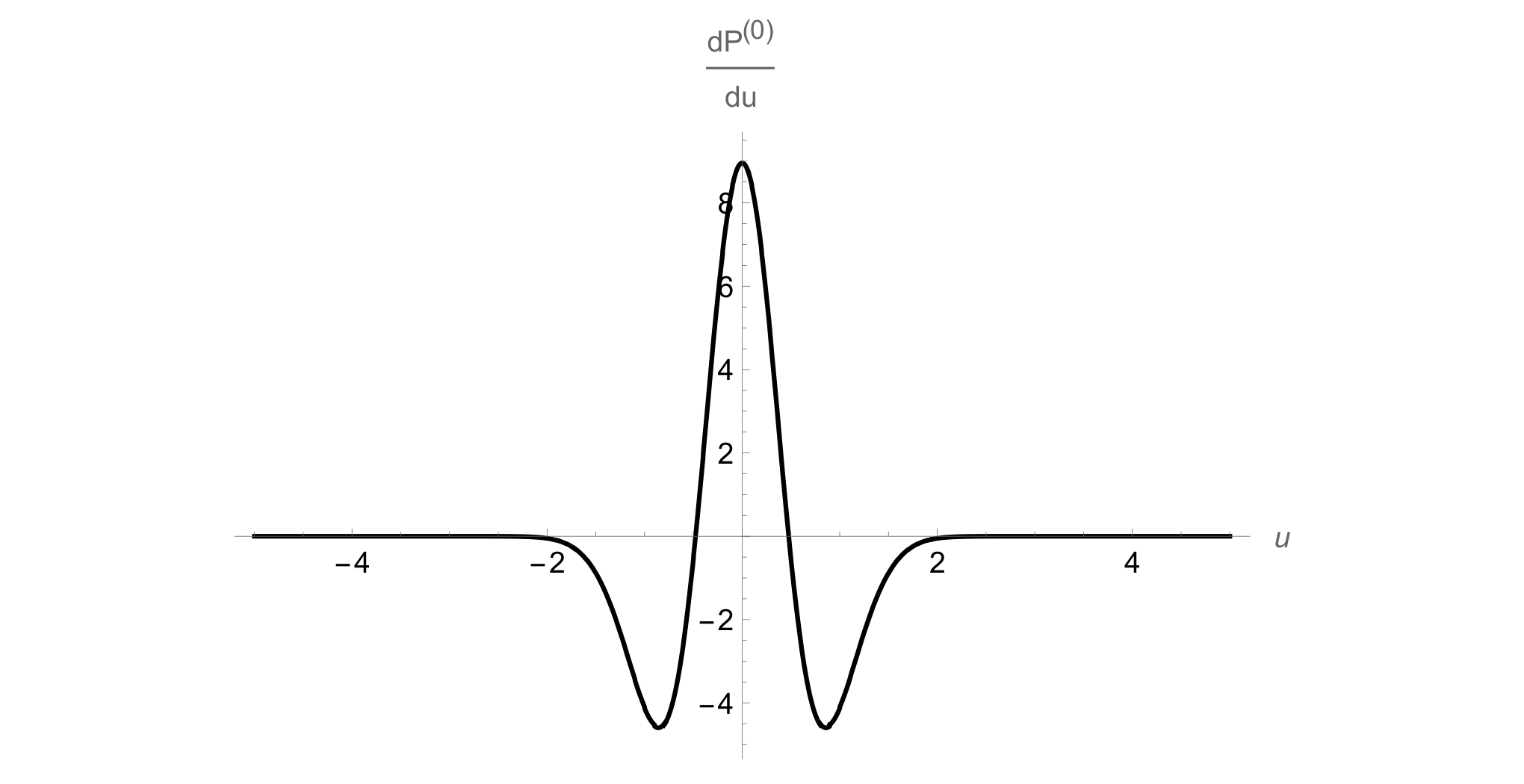}
\par\smallskip
{\small (b) Energy flux $\mathrm{d}P^{(0)}/\mathrm{d}u$.}
\end{minipage}
\caption{General radiative configuration expanded in spherical harmonics with
$\ell_{\min}=2$, $\ell_{\max}=3$, and initial mass aspect $M_0(\theta,\phi)=1$.}
\label{fig:bondi_general_energy_flux}
\end{figure}

For a given initial Bondi mass aspect $M_0(\theta,\phi)=1$ and multipoles in the range
$\ell_{\min}=2$ to $\ell_{\max}=3$, the total gravitational energy $P^{(0)}(u)$ and its retarded-time
derivative are obtained by numerically integrating the corresponding expressions over the unit
sphere. The resulting energy and energy flux are shown in Fig.~\ref{fig:bondi_general_energy_flux}. They show two remarkable and closely related features. First, the total gravitational energy exhibits a net negative variation, indicating that the radiative process leads to a negative contribution to the asymptotic energy balance. Second, the energy flux presents a pronounced peak around $u=0$, which corresponds to the passage of the gravitational-wave front across null infinity. This behavior is a direct consequence of the localized temporal support of the radiative modes and provides a clear identification of the wavefront in terms of the retarded time. An important implication of the negative character of the gravitational energy is that, unlike ordinary radiative fields, it cannot be dissipated through interactions with matter during propagation in the usual attenuating sense. In conventional situations, energy loss due to interaction leads to attenuation of the wave. In the present case, however, any reduction in the energy carried by the radiation makes its contribution more negative, thereby reinforcing rather than suppressing its effect. This distinctive property allows gravitational radiation with negative energy to persist and accumulate over large scales, providing a natural mechanism through which radiative processes may build up in regions approaching the cosmological horizon.

Although the general spherical-harmonic expansion provides a convenient framework to explore
radiative configurations, it is important to verify whether the features identified above persist
in the original Bondi--Sachs setting. This is achieved by considering the axially symmetric case,
in which the radiative degrees of freedom reduce to a single shear function $c(u,\theta)$ and the
second polarization is suppressed by setting $d(u,\theta)=0$. In this situation the space-time
corresponds precisely to the standard Bondi radiative configuration, with no azimuthal dependence
and no additional angular structure beyond that originally introduced by Bondi.

\begin{figure}[t]
\centering
\begin{minipage}{0.48\linewidth}
\centering
\includegraphics[width=\linewidth]{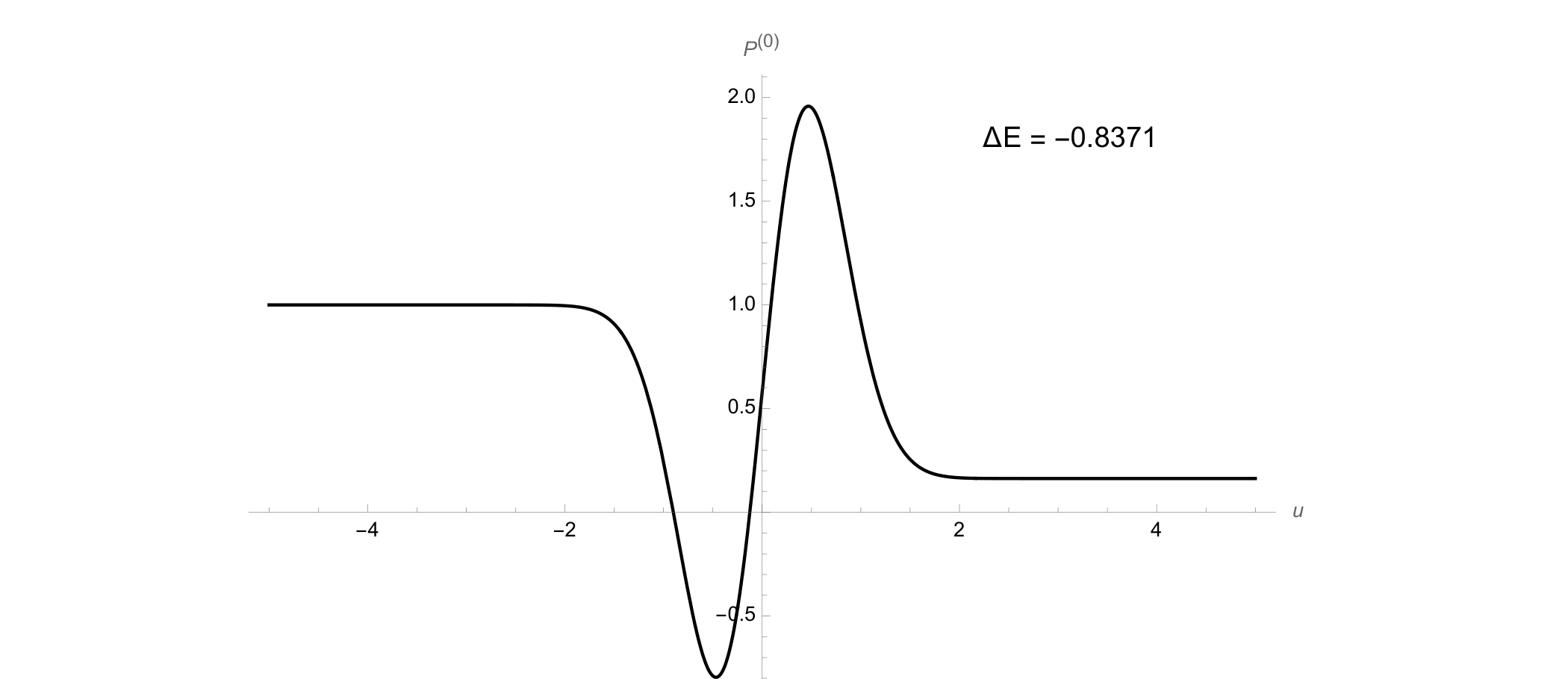}
\par\smallskip
{\small (a) Total Bondi gravitational energy $P^{(0)}(u)$ in the axially symmetric configuration.}
\end{minipage}\hfill
\begin{minipage}{0.48\linewidth}
\centering
\includegraphics[width=\linewidth]{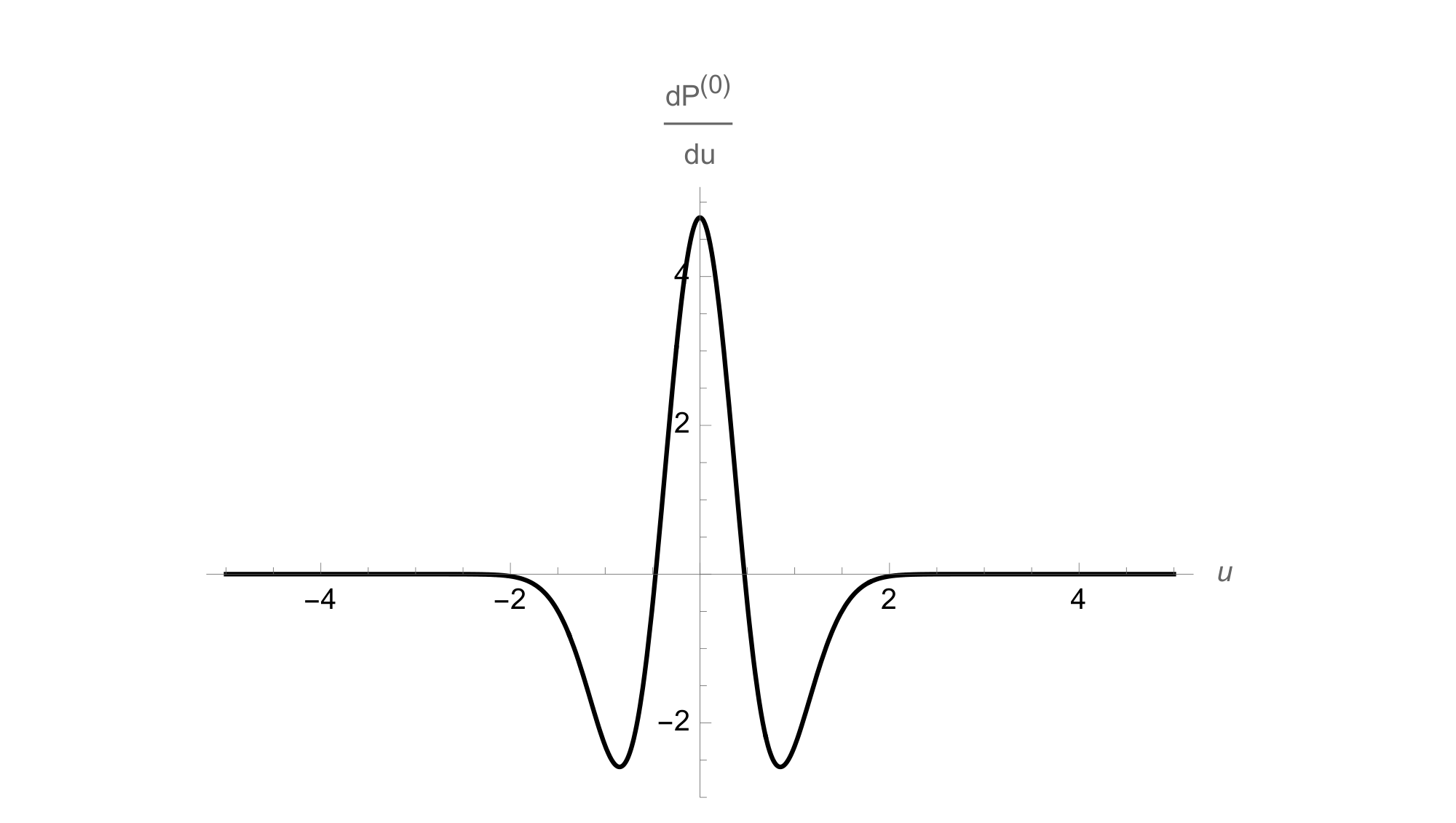}
\par\smallskip
{\small (b) Energy flux $\mathrm{d}P^{(0)}/\mathrm{d}u$ associated with the axially symmetric radiation.}
\end{minipage}
\caption{Axially symmetric Bondi radiative configuration with $d=0$,
$\ell_{\min}=2$, $\ell_{\max}=4$, and initial mass aspect $M_{0}(\theta)=1$.}
\label{fig:bondi_axial_energy_flux}
\end{figure}

In the axially symmetric case, the radiative degrees of freedom are specified by setting
$d(u,\theta)=0$ and prescribing the shear function $c(u,\theta)$ in terms of Legendre polynomials,
\begin{equation}
c(u,\theta) =
\left(1-\cos^{2}\theta\right)
\sum_{n=\ell_{\min}}^{\ell_{\max}}
\frac{e^{-u^{2}}}{n^{2}}
\,\frac{d^{2}}{dx^{2}}P_{n}(x)
\bigg|_{x=\cos\theta},
\end{equation}
which corresponds to the standard Bondi radiative structure with no azimuthal dependence. For this configuration we again compute the total gravitational energy $P^{(0)}(u)$ and the
associated energy flux $\mathrm{d}P^{(0)}/\mathrm{d}u$ by numerically integrating the corresponding
expressions over the sphere. The calculation is performed for multipoles in the range
$\ell_{\min}=2$ to $\ell_{\max}=4$, assuming an initial Bondi mass aspect $M_{0}(\theta)=1$.
The resulting energy and energy flux are shown in Fig.~\ref{fig:bondi_axial_energy_flux}. The results indicate that the qualitative behavior of the total gravitational energy and its associated flux remains unchanged when compared with the general radiative configuration. In particular, the net negative variation of the energy and the localization of the flux around the wavefront persist in the axially symmetric Bondi case. The restriction to $d=0$ and the elimination of azimuthal dependence affect only the overall intensity of the effect, while its fundamental features, namely the sign of the energy transfer and its temporal localization, are preserved. This demonstrates that the mechanism identified here is not sensitive to the angular complexity of the radiation, but is an intrinsic property of the Bondi radiative process itself.

Having established that the qualitative features of the energy balance remain unchanged in the
absence of azimuthal dependence, we now turn to another genuine aspect of the Bondi radiative
process: the transfer of linear momentum. In the axially symmetric configuration with
$d(u,\theta)=0$, the space-time corresponds exactly to the original Bondi--Sachs solution, and no
additional degrees of freedom are introduced. In this context, any variation of the spatial
components of the total energy--momentum vector arises solely from the anisotropic emission of
gravitational radiation inherent to the Bondi mechanism itself. For this genuine Bondi configuration, we compute the longitudinal component of the Bondi momentum,
$P^{(3)}(u)$, and define the associated radiative force as
\begin{equation}
F^{(3)}(u) \equiv \frac{dP^{(3)}(u)}{du}.
\end{equation}
This quantity measures the rate at which linear momentum is carried away by gravitational
radiation across null infinity.

\begin{figure}[t]
\centering
\begin{minipage}{0.48\linewidth}
\centering
\includegraphics[width=\linewidth]{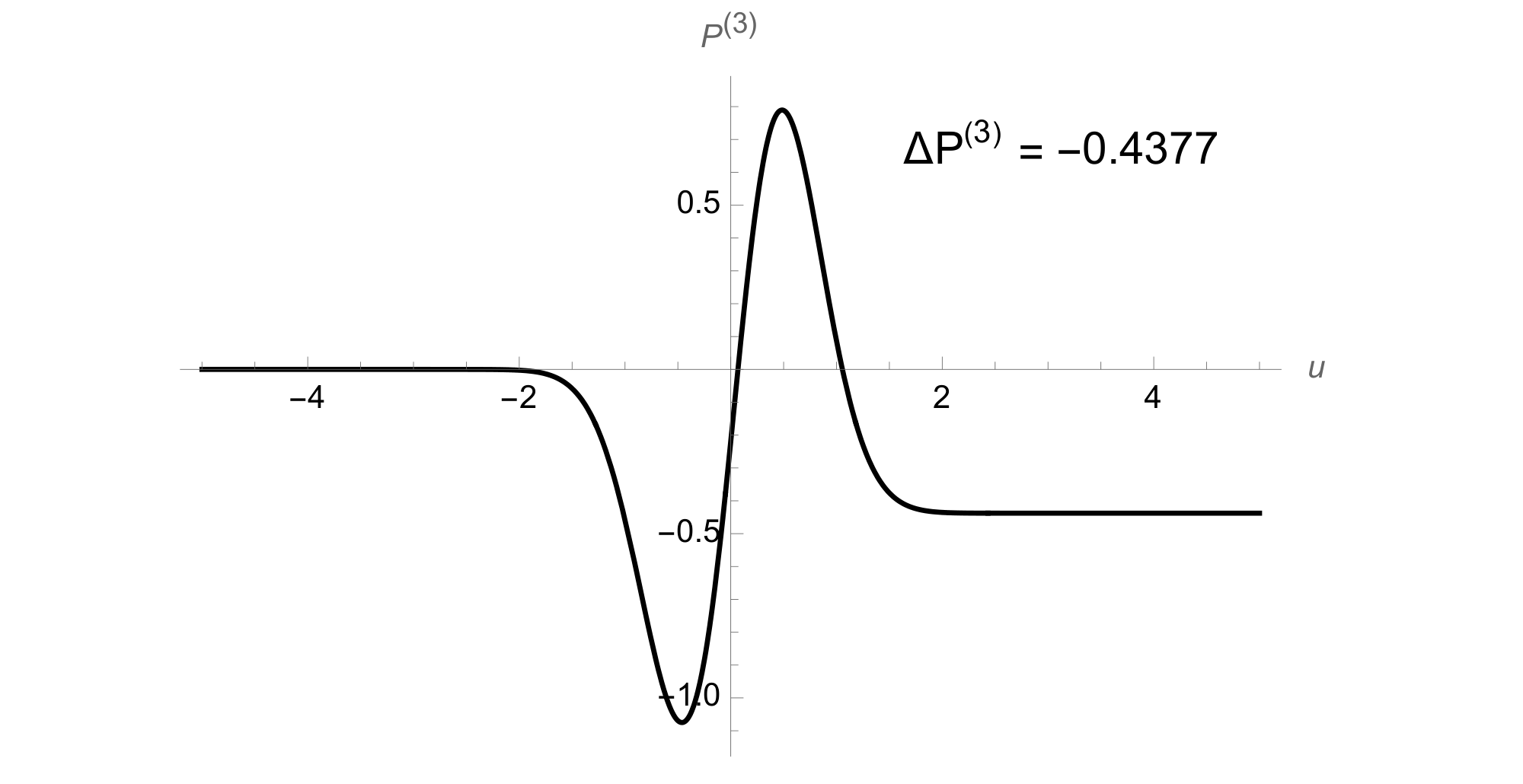}
\par\smallskip
{\small (a) Bondi momentum component $P^{(3)}(u)$ as a function of the retarded time.}
\end{minipage}\hfill
\begin{minipage}{0.48\linewidth}
\centering
\includegraphics[width=\linewidth]{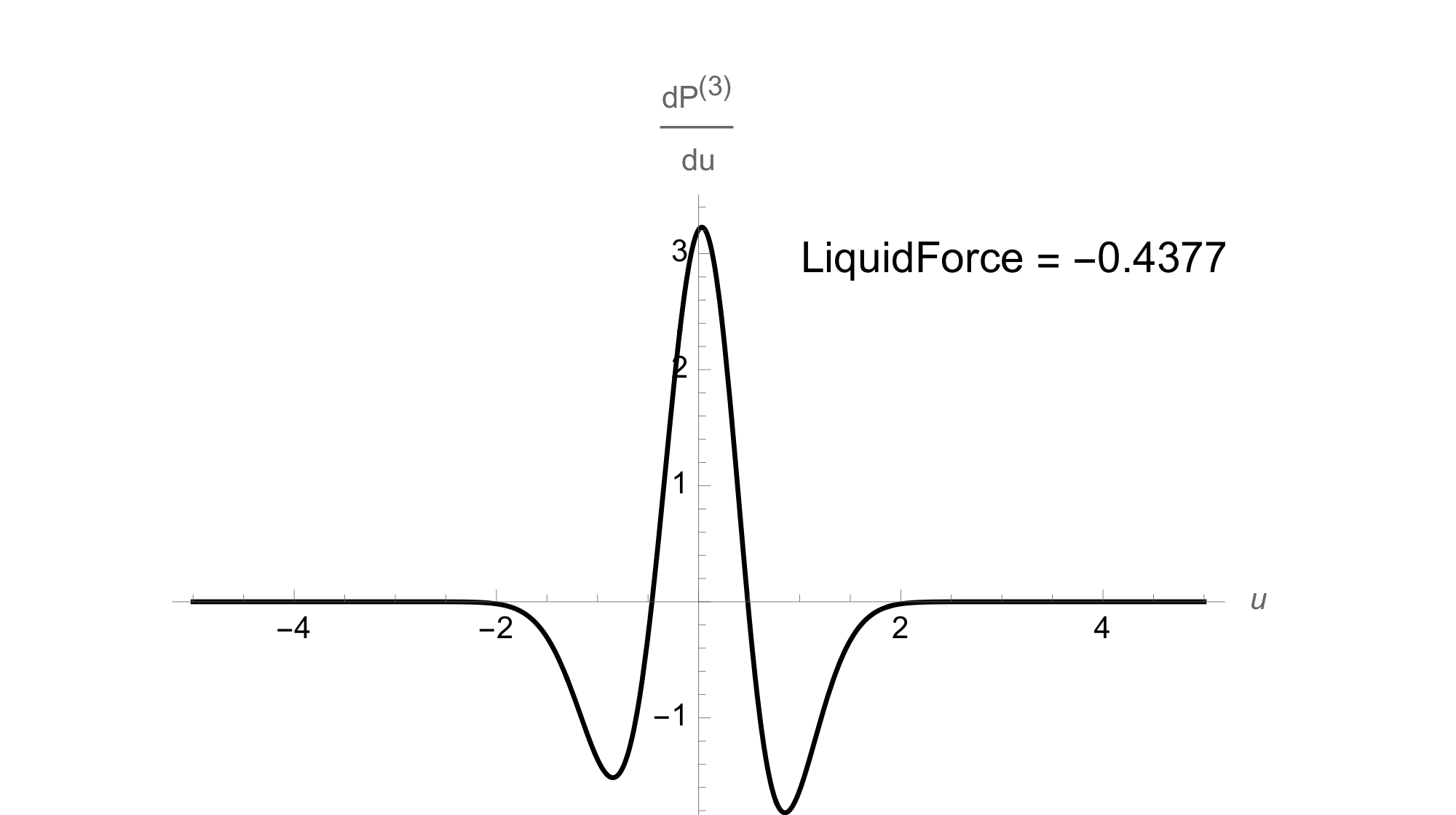}
\par\smallskip
{\small (b) Radiative force $F^{(3)}(u)=dP^{(3)}/du$.}
\end{minipage}
\caption{Bondi momentum transfer in the genuine Bondi--Sachs radiative configuration
($d=0$), with $\ell_{\min}=2$, $\ell_{\max}=4$, and initial mass aspect $M_{0}=1$.}
\label{fig:bondi_momentum_force}
\end{figure}

The momentum component $P^{(3)}(u)$ and the corresponding radiative force $F^{(3)}(u)$ are obtained
by numerical integration over the sphere using the same Bondi parameters adopted in the previous
analysis. The components $P^{(1)}(u)$ and $P^{(2)}(u)$ are zero for the axially symmetric Bondi spacetime. The resulting momentum variation and force profile are shown in
Fig.~\ref{fig:bondi_momentum_force}.

\section{Conclusion} \label{sec.4}

In this work we have investigated the role of gravitational energy and its associated flux in
radiative Bondi--Sachs space-times within the teleparallel equivalent of general relativity.
The analysis was based on expressions for the total gravitational energy and energy--momentum
flux previously obtained in the literature, which have been successfully employed, for instance,
to characterize the contribution of gravitational radiation as an effective noise source in
gravitational-wave detectors. Here, rather than revisiting their derivation, we have explored a
different physical regime and provided a new interpretation of these quantities in a cosmological
context.

By explicitly evolving the Bondi mass aspect through the Bondi--Sachs evolution equation and
analyzing representative radiative configurations, we have shown that gravitational radiation
leads to a net negative contribution to the total gravitational energy in asymptotic regions.
The associated energy flux is strongly localized around the passage of the wavefront, yet its
integrated effect remains finite and negative. An important consequence of this behavior is that
gravitational radiation carrying negative energy cannot dissipate through interaction with matter
during propagation. Any reduction of the energy carried by the radiation makes its contribution
more negative, allowing the radiative energy to persist and accumulate over large distances and
long time scales.

The analysis of the momentum sector further reveals that radiative Bondi--Sachs space-times
exhibit a nonvanishing transfer of linear momentum. The retarded-time derivative of the Bondi
momentum defines an effective radiative force, whose integral yields a net negative impulse.
For a single radiative source, this force is localized near the wavefront and corresponds to an
effective attraction toward the region where the gravitational radiation concentrates. This result
shows that a single Bondi radiative process is already sufficient to induce a directed dynamical
effect on the surrounding space-time.

Taken together, the behavior of the energy, flux, momentum, and force suggests a consistent
physical picture. Gravitational radiation produced throughout cosmic history may accumulate in
regions approaching the causal boundary of the universe, where its negative energy contribution
and associated momentum transfer act as an effective large-scale dynamical agent. Although each
individual radiative event produces a small and localized effect, the cumulative contribution of
many such processes may influence the motion, and perhaps the acceleration, of matter within the causal region. The mechanism analyzed here provides a contribution to cosmic acceleration driven by the cumulative effect of gravitational radiation energy and momentum generated by
radiative mass--energy processes. The mechanism discussed here is local and does not rely on a specific cosmological background, although its cumulative effects naturally acquire significance when considered over cosmological time scales and within the observable horizon.

We also observe that the Bondi--Sachs mechanism induces a permanent modification in the energy configuration of spacetime, i.e., the total gravitational energy evaluated at null infinity exhibits a net decrease, signaling an irreversible energy loss due to gravitational radiation. This behavior closely resembles a memory effect, not associated with the dynamics of test particles as in the conventional gravitational wave memory but rather with the spacetime energy itself. In this sense, the effect may be interpreted as an energy memory of the spacetime geometry. Notably, when the mass aspect is suppressed, this permanent energy shift disappears, and the energy evolution reverts to the pattern previously reported in Ref.~\cite{ulhoa2024}.

A natural continuation of the present investigation is to incorporate the gravitational radiation produced by the Bondi mechanism as an effective fluid in the Einstein equations, using the equation of state  established in Ref.~\cite{Ulhoa2024MPLA}. This would connect the present mechanism to the Friedmann framework and allow for a quantitative assessment of its contribution to cosmic acceleration.

\bibliographystyle{apsrev}
\bibliography{refs}

\end{document}